\documentclass{PoS}
\usepackage{epsfig}

\title{Eight powers of ten: similarities in black hole accretion on
all mass scales}

\ShortTitle{Black hole accretion}

\author{\speaker{Rob Fender}\\
        University of Southampton, UK\\
        E-mail: \email{rpf@phys.soton.ac.uk}}
\author{Elmar K\"ording\\
        University of Southampton, UK\\}
\author{Tomaso Belloni\\
        Osservatorio Astronomico di Brera, Italy\\}
\author{Phil Uttley\\
        University of Amsterdam, NL\\}
\author{Ian McHardy\\
        University of Southampton, UK\\}
\author{Tasso Tzioumis\\
        ATNF, Australia\\}

\abstract{ In this paper we discuss the recent advances in the
quantitative comparison of accretion, and the accretion:jet coupling,
in accreting black holes in both X-ray binaries (where $M_{\rm BH}
\sim 10M_{\odot}$) and Active Galactic Nuclei ($10^5 \leq M_{\rm BH}
\leq 10^9$). These similarities include the radiative efficiency and
jet power as a function of accretion rate, which are themselves
probably the origin of the `fundamental plane of black hole
activity'. A second `fundamental plane' which connects mass, accretion
rate and timing properties provides us with a further physical
diagnostic. Patterns of radio loudness (i.e. jet production) as a
function of luminosity and accretion state are shown to be similar for
X-ray binaries and AGN. Finally we discuss how neutron stars are a
useful control sample, and what the future may hold for this field.}

\FullConference{VI Microquasar Workshop: Microquasars and Beyond\\
		September 18-22 2006\\
		Societ\`a del Casino, Como, Italy}

\begin{document}

\section{Similarities between XRB and AGN: not a new idea...}

Black holes are extremely simple objects which we find at large in our
universe over an enormous range (at least eight orders of magnitude)
of physical scales. At both extremes of this scale their most obvious
property is their copious power output as a result of accretion. In
this paper some recent results on the similarity of this accretion
process, and the related outflows, across this extreme range of scale,
will be discussed. Before we do so it is important however to note that
the concept that black hole accretion might in some sense be
scale-invariant, or only weakly dependent on scale, is not new. 

In 1976, Shakura \& Sunyaev clearly outlined what they expected to be
the simple scaling laws between stellar mass black holes in X-ray
binary systems, and supermassive black holes in quasars. McHardy
(1988) noted similarities in the power spectra of both classes of
system, and Sams, Eckart \& Sunyaev (1996) discussed the scaling of
black hole jets with mass (a discussion expanded upon in some detail
by Heinz \& Sunyaev (2003)). Falcke \& Biermann (1996) further
elaborated in some detail on the likely accretion:jet coupling in
black holes on all scales, while Pounds, Done \& Osborne (1995)
discussed the similarities between Seyfert X-ray emission and that of
black hole X-ray binaries in soft X-ray states. Famously, Mirabel et
al. (1992) and Mirabel \& Rodriguez (1994, 1999) coined the phrase
`microquasar' for jet-producing X-ray binaries, inescapably fixing the
analogy in the minds of all researchers. Markoff et al. (2001, 2003)
have taken models inspired by low-luminosity AGN and adapted them to
the hard state of black hole X-ray binaries.  Most recently,
Maccarone, Fender \& Ho (2005) comprises a compilation of papers on
this subject.

\subsection*{Why we expect black hole accretion to be essentially scale-free}

Extreme mathematical simplicity !
For all black holes, the ratio between the mass and radius of the
event horizon, $M / R$, is the same, to within a factor of 

\begin{itemize}
\item{{\bf two} if we consider R at the event horizon}
\item{{\bf six} if we consider R at the Innermost Stable Circular Orbit (ISCO)}
\end{itemize}

-- in both cases the largest R corresponds to a non-spinning
   (Schwarzschild) black hole, and the smallest R to a maximally
   rotating (`Maximal Kerr') black hole.

There are no other parameters (besides charge, which is generally
considered to be unimportant), i.e. black holes have no hair
(e.g. Carter 1973). Characteristic timescales such as the orbital
period at the innermost stable circular orbit likewise scale linearly
with mass. Observations indicate an observed range of masses, and
therefore of physical scales, of about $10^8$ over which this scaling
should hold. No other objects in the universe scale so perfectly.

\subsection*{Why we do not expect black hole accretion to be essentially scale-free}

Microphysics ! The matter at the inner edge (or elsewhere) of an X-ray
binary accretion disc is much hotter and probably much denser than
that in an accretion disc around a supermassive black hole (and who
knows about magnetic field ?).

\section{The fundamental plane of black hole activity}

\begin{figure}
\centerline{\epsfig{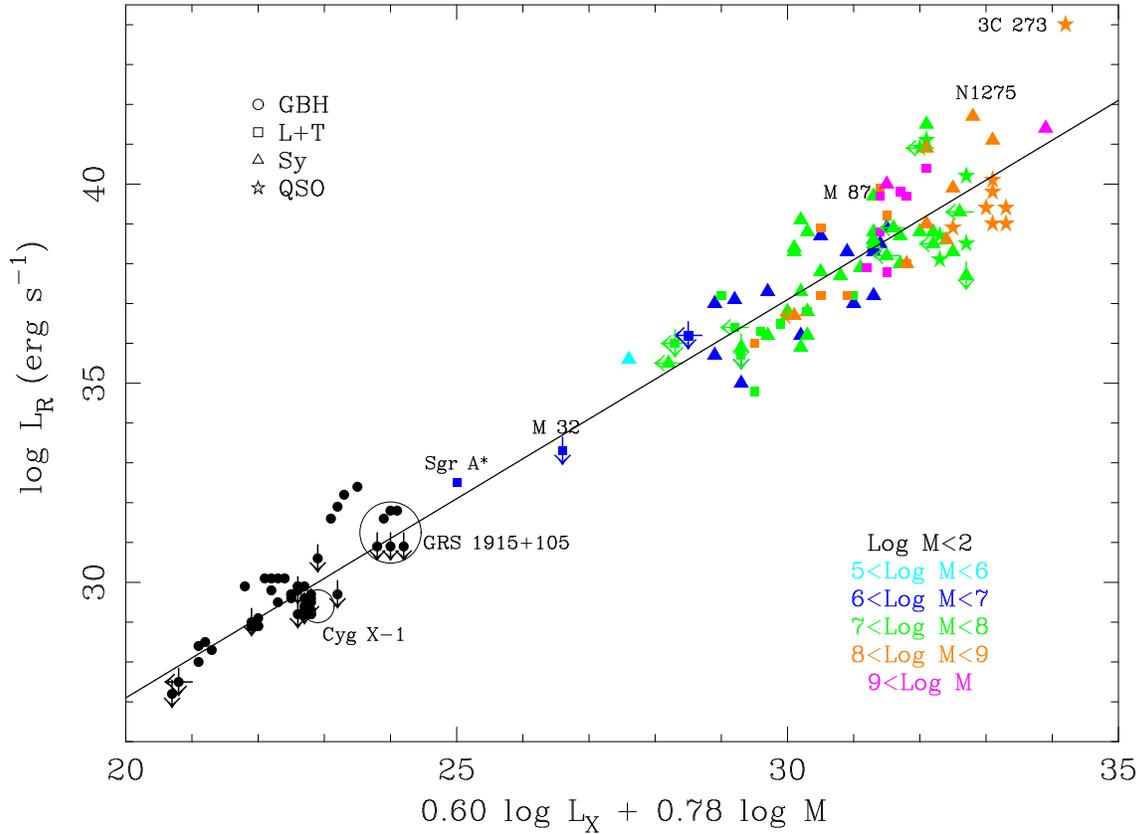}}
\label{fp1}
\caption{The fundamental plane of black hole activity, relating radio
luminosity, X-ray luminosity and mass over more than eight orders of
magnitude in black hole mass. A major step forward in the unification
of accreting black holes on all mass scales. From Merloni, Heinz \& di
Matteo (2003); Falcke, K\"ording \& Markoff (2004) independently
arrived at almost the same result.}
\end{figure}

Corbel et al. (2000, 2003) demonstrated a clear nonlinear relation between
the radio luminosty $L_{\rm radio}$ and X-ray luminosity, $L_{\rm X}$
in the black hole binary GX 339-4 (this source is a fantastic testbed
for such studies since it regularly varies by 4--5 order of magnitude
in $L_X$ and has no detectable / contaminating companion star), of the form

\[
L_{\rm radio} \propto L_X^{0.7}
\]

Gallo, Fender \& Pooley (2003), compiled a larger sample of more black
hole binaries and found this relation to be essentially universal, an
assertion recently confirmed by the detection of radio emission from
the quiescent ($L_X \sim 10^{-8} L_{\rm Edd}$) black hole binary A
0620-00 at the level predicted by the relation (Gallo et al. 2006;
note the exponent of the correlation flattens slightly to +0.6 when
this quiescent source is included). However, it should be noted that
an increasing number of sources are being found at relatively high
luminosities ($L_X \geq 10^{-3} L_{\rm Edd}$) which drop off the
correlation (Gallo 2007). Some possible explanations for this effect
could include partial quenching or increased velocity of the jets as
the soft state is approached, but more data are required.

Shortly afterwards two groups (Merloni, Heinz \& di Matteo 2003;
Falcke, K\"ording \& Markoff 2004) independently established the
existence of a plane linking $L_{\rm radio}$, $L_X$ and mass $M$ of
all accreting black holes, from X-ray binaries to AGN, and including
Sgr A* (Fig 1). This should be considered one of the major
steps in the unification of black hole accretion on all mass scales.

In the Merloni et al. formalism, the fundamental plane can be
represented as:

\[
L_{\rm radio} \propto L_X^{0.6} M^{0.8}
\]

where the power-law indices are fitted values to a large sample of
XRBs and AGN.

The most recent refinements of the plane are presented in Gallo et
al. (2006), K\"ording, Falcke \& Corbel (2006). Criticisms of the
plane have been rebuked by a consortium of all the original discovery
authors, in Merloni et al. (2006).

\subsection{Calibrating the plane}

\begin{figure}
\centerline{\epsfig{file=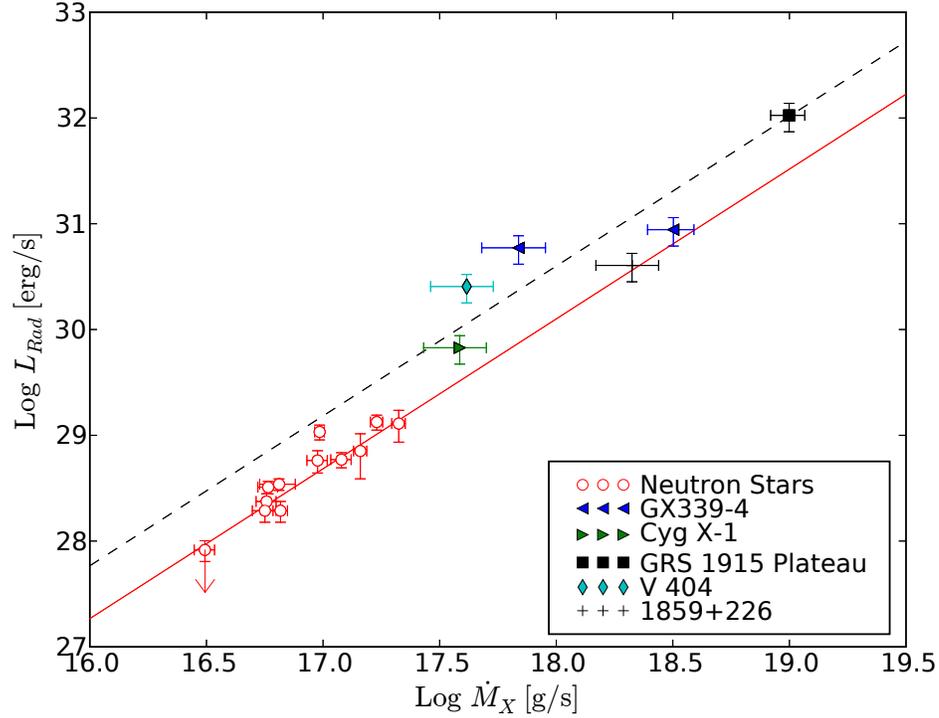, angle=0, width=14cm}}
\caption{The correlation between accretion rate and radio luminosity
for a sample of black hole and neutron star X-ray binaries. The lines
indicate fits with gradient +1.4, which is expected for the condition
that jet power is linearly proportional to accretion rate. From
K\"ording, Fender \& Migliari (2006).}
\end{figure}

Of the three parameters of the fundamental plane, one is genuinely
fundamental ($M$), and one is a good indication of the total radiative
output of the system and is therefore pretty fundamental ($L_X$).
However, the third parameter, $L_{\rm radio}$ is merely a tiny tracer
of the enormous power carried by the jets from these systems. The fact
that it seems to correlate so perfectly with $L_X$ is itself quite
amazing and indicates a remarkable stability and regularity in the jet
formation process.

The plane would be a much better indicator of physical quantities and
the flow of matter and power around the black hole if $L_{\rm radio}$
could be replaced with, say, the total jet power $L_J$ or the mass
accretion rate $\dot{m}$. In fact we can do both, based upon the
relations established in K\"ording, Fender \& Migliari (2006;
hereafter KFM03; see also e.g. Heinz \& Grimm 2005). In Fig 2 we
present the relation between accretion rate (as determined from X-rays
in radiatively efficient states) and radio luminosity from K\"ording,
Fender \& Migliari (2006). In Fig 3 we use this and the related
relation between jet power and radio luminosity from the same paper,
to `calibrate' the plane, by replacing the axes of Merloni et al. with
the physical quantities of jet power (left) and accretion rate
(right).

\begin{figure}
\centerline{\epsfig{file=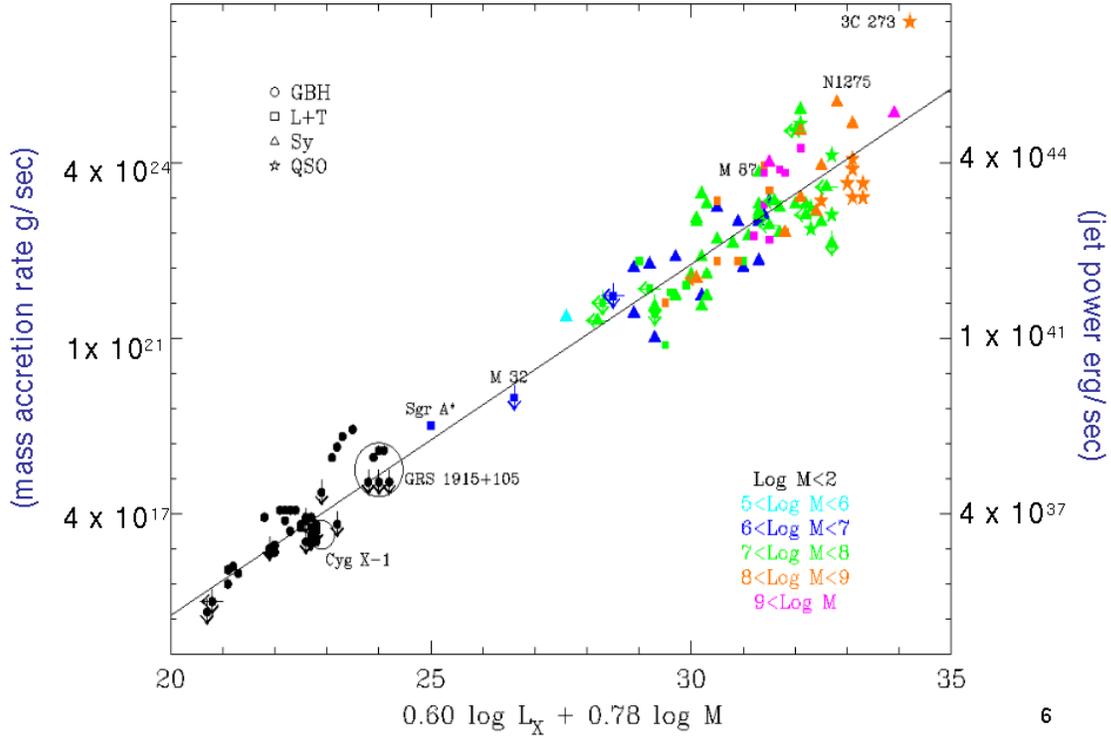, angle=0, width=15cm}}
\label{fp2}
\caption{The `calibrated' fundamental plane, where radio luminosity
can be replaced by total jet power and/or mass accretion rate (see Fig
2), which are more physically useful numbers.}
\end{figure}

\subsection{What does the plane mean ?}

The fundamental plane is an observational fact, a fit to measured
quantities. But what does it mean ? In the following we outline what
is probably the most straightforward interpretation.

If 

\[
L_{\rm radio} \propto \dot{m}^{1.4}
\]

which we consider to have been demonstrated in KFM03 (see our Fig 2), and

\[
L_X / L_{\rm Edd} \propto (\dot{m} / \dot{m}_{\rm Edd})^2
\]

which is a general approximate solution for radiatively inefficient
accretion (where the accretion flow knows which Eddington ratio it is
at), then simple rearranging gives us:

\[
L_{\rm radio} \propto L_X^{0.7} M^{0.7}
\]

which is within $\pm 0.1$ in power-law indices of the fitted plane.

This is not the only possible explanation for the existence of the
fundamental plane, but it is probably the simplest. However, for it to
work as an explanation, it requires that the fundamental plane be
dominated by {\em radiatively inefficient} sources. Can that be right
?  K\"ording, Fender \& Migliari (2006) argue that this is indeed the
case (note that Merloni, Heinz \& di Matteo (2005) discuss this
interpretation in a very similar way).

\begin{figure}
\centerline{\epsfig{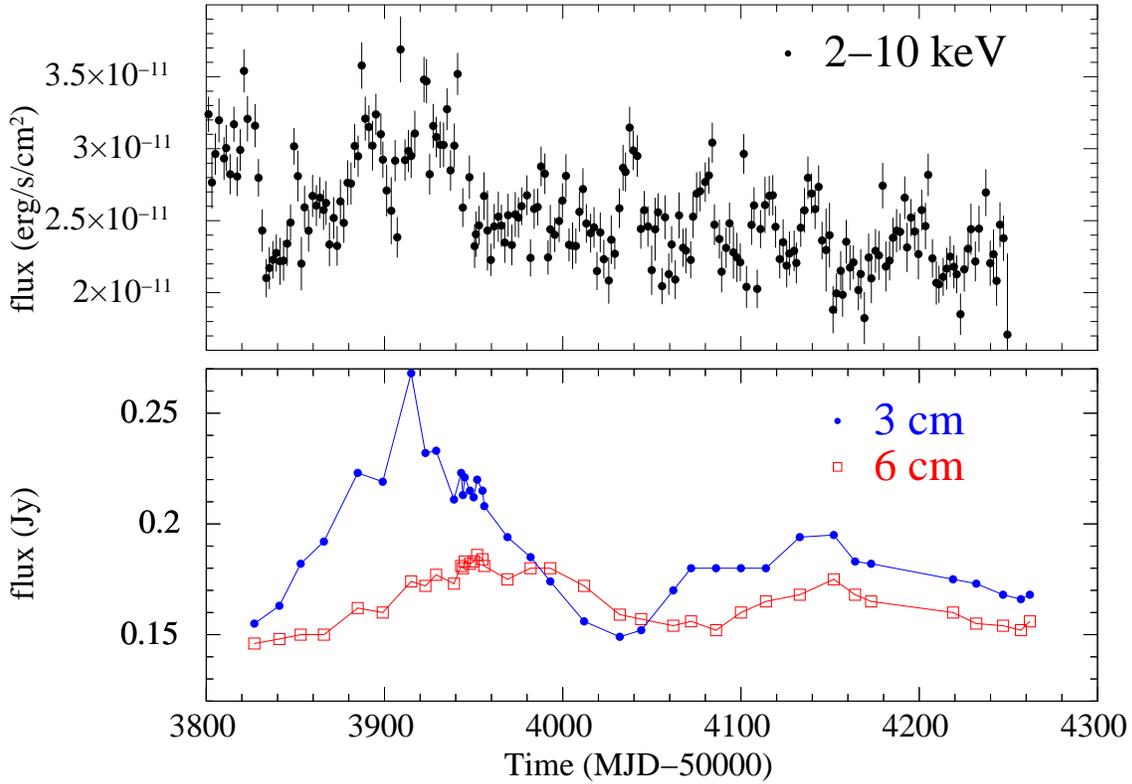}}
\label{plot-corr}
\caption{X-ray (RXTE) and radio (ATCA) monitoring of the
low-luminosity AGN NGC 7213, for which monitoring continues. The goal
of these observations is to look for lags and correlations between the
two bands to try and recover the `intrinsic' correlation which for AGN
is so far only implied by the global effect (unlike for X-ray
binaries, in which both are seen). Monitoring continues.}
\end{figure}

\section{Searching for the radio:X-ray correlation in a single AGN}

It is notable that in X-ray binaries both `global' and `intrinsic'
correlations have been found, by which we mean that a range of points
from a variety of sources traces out the same correlation as is
observed from two individual sources (GX 339-4 and V404 Cyg). In the
case of AGN, only the `global' effect has been observed (although note
the possible cyclic disc-jet coupling in 3C120 (Marscher et
al. 2002)). Could it be possible to see something like the $L_{\rm radio}
\propto L_{\rm X}^{0.7}$ relation in a single AGN ?

We have been attempting to do this with a program of X-ray (RXTE) and
radio (ATCA) monitoring of the highly variable AGN NGC 7213 (see Fig
4). To date we have monitored the source for nearly a year and a half
and see both the expected lags between the radio bands and some hints
of correlation between radio and X-rays. Monitoring continues.

\section{Timing properties: another plane}

\begin{figure}
\centerline{\epsfig{file=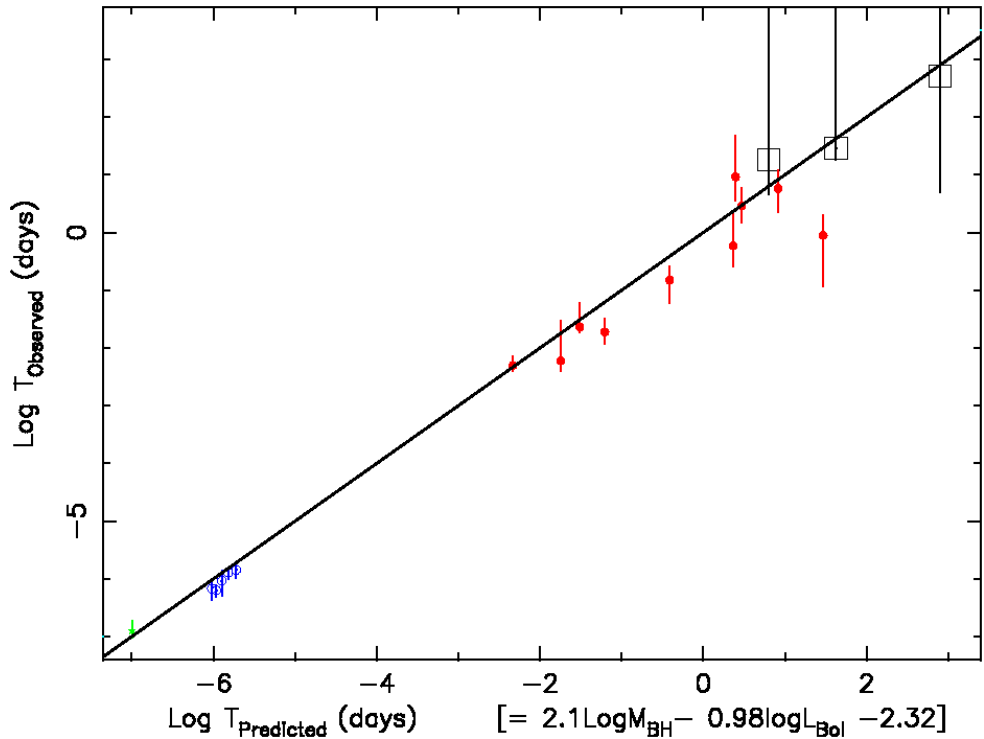, angle=0, width=13cm}}
\caption{A second fundamental plane, essentially relating
characteristic timescales to mass and accretion rate, from McHardy et
al. (2006). More recently K\"ording et al. (2007) have shown this to
extend `hard state' black hole X-ray binaries {\em} and neutron star
systems.}
\end{figure}

AGN X-ray timing studies started in the 1980s with the launch of
EXOSAT, which probed AGN X-ray variability on time-scales up to a few
days.  It was soon established that on these time-scales AGN showed
red-noise variability with steep unbroken PSDs (Lawrence et al. 1987,
McHardy et al. 1987).  However, McHardy (1988) noted using sparse
long-term archival data that the PSDs appeared to flatten or break on
longer time-scales, similar to the PSD shapes seen in BH XRBs.  The
launch of RXTE in 1995 allowed long-term light curves to be obtained
with extremely good sampling and categorically proved the existence of
PSD breaks on time-scales close to those expected by scaling the BH
XRB break time-scales up by the AGN BH mass (e.g. Uttley et al. 2002,
2005; Markowitz et al. 2003, McHardy et al. 2004), although with some
considerable scatter in the mass-time-scale relation. 

Recently, Migliari, Fender \& van der Klis (2005) established that for
a small sample of X-ray binaries there was a positive correlation
between radio luminosity and the frequencies of timing features. This
was to be expected, since we are confident that on the whole both
timing frequencies and radio luminosity are increasing functions of
accretion rate (although we also know there are other, state-related,
dependencies at high accretion rates). 

We (McHardy et al. 2006) have now fitted a plane which relates mass
and accretion rate to the break frequency in X-ray power spectra, such
that

\[
T_{\rm break} \propto M^{-2.1} L_{\rm bol}^{-1}
\]

Where $T_{\rm break}$ is the break timescale, reciprocal of the break
frequency, $M$ is black hole mass and $L_{\rm bol}$ is bolometric
luminosity. All of the sources used in the correlation are believed to
be in radiatively efficient states and so $L_{\rm bol}$ is used as a
proxy for accretion rate. Using this substitution converted to
accretion rate and integer power-law indices (i.e. $2 \sim 2.1$), we
arrive at

\[
T_{\rm break} \propto M / (\dot{m} / \dot{m}_{\rm Edd})
\]

revealing the expected linear correlation of break timescales with
black hole mass, albeit for a fixed Eddington ratio of accretion
rate. Our next goal is to see how well we can extend this relation to
other accretion states and perhaps even other classes of object
(K\"ording et al. 2007).

\section{Similarities in disc-jet coupling}

Black hole X-ray binaries seem to (nearly) always follow a pattern of
behaviour in outburst similar to that sketched in Fender, Belloni \&
Gallo (2004) and Homan \& Belloni (2005) (see Figs 6 and 7). However,
it is clear that between different outbursts of the same source, or
outbursts of different sources, the luminosities at which the hard
$\rightarrow$ soft and soft $\rightarrow$ hard state transitions may
occur can vary quite significantly (e.g. Belloni et al. 2006). As a
result, an ensemble of X-ray binaries would present a pattern in the
hardness:luminosity diagram which resembled a golf club {\footnote{or
a turtle's head, as was suggested in a bar at the Texas Symposium in
2004..}}, with a long handle and a filled-in head. Such an ensemble is
obviously what we're going to have to deal with if we want to be able
to compare patterns of disc:jet coupling in XRBs with those in AGN.

In K\"ording, Jester \& Fender (2006) we have attempted to do
this. First we constructed the Disc Fraction - Luminosity Diagram, in
which hardness is replaced by the ratio of power law to total
luminosity, a number which approaches zero for disc dominated soft
states, and unity for hard states. This is necessary for a physical
comparison, since the accretion discs temperature is a decreasing
function of black hole mass, and for AGN does not contribute
significantly in the X-ray band. We then simulated an ensemble of
BHXRBs, based upon Fender, Belloni \& Gallo (2004) and the slight
refinement (suggested in Belloni et al. 2006) that the `jet line'
might be diagonal in such a diagram. This was then compared to a
sample of AGN from the SDSS DR5 for which there were X-ray detections
and either radio detections or limits, {\em plus} a sample of
low-luminosity AGN (LLAGN). The similarity was striking (see Fig 6)
and suggests that the radio loudness is determined by the combination
of `state' and luminosity in a similar way for accreting black holes
of all masses. Note that, while it is tempting to consider, the
diagram does {\em not} indicate that AGN necessarily follow the same
anti-clockwise loop in the diagram as XRBs: the motion in such loops
could possibly be dominated by disc instabilities which may not apply
to AGN. What it does indicate is that when an AGN finds itself in a
particular accretion `state', whether disc or corona dominated or some
mix of the two, the jet it produces will be comparable to that which a
XRB would make in the same state.

\begin{figure}
\centerline{\epsfig{file=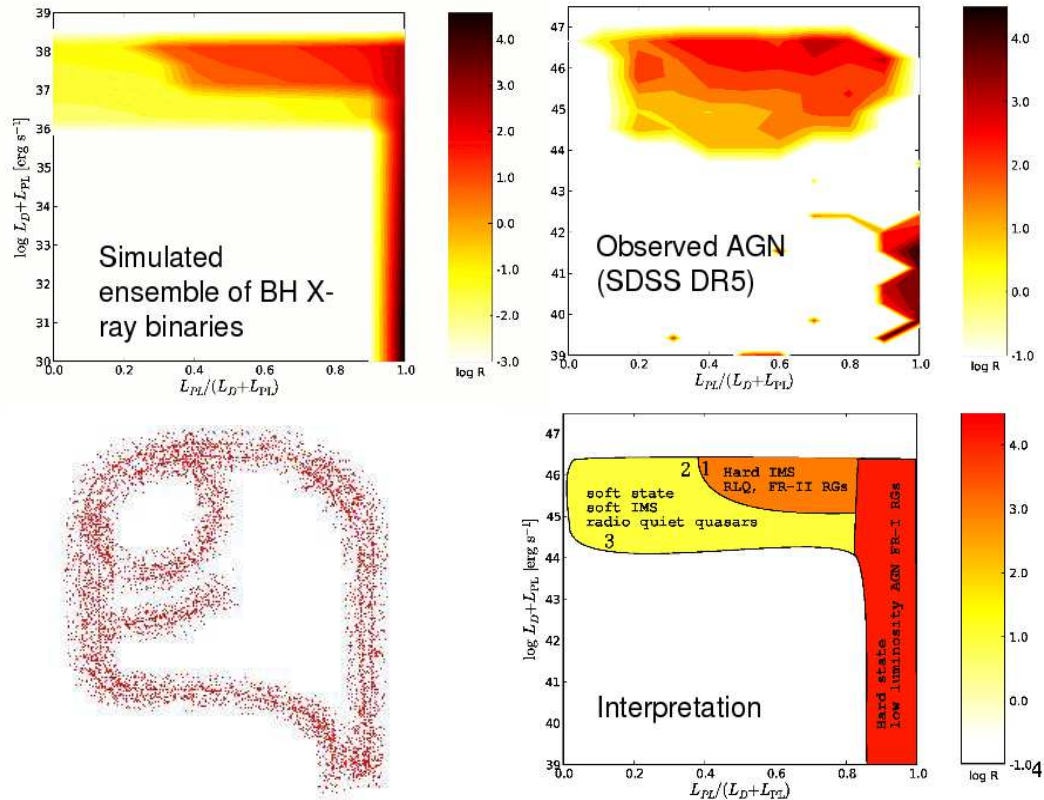, angle=0, width=15cm}}
\caption{A comparison of the disc:jet coupling in X-ray binaries with
that in AGN, based upon the Disc Fraction Luminosity Diagram (DFLD;
see K\"ording, Jester \& Fender 2006). Based upon the disc:jet model
for black hole X-ray binaries presented in Fender, Belloni \& Gallo
(2004), an example track for which is shown in the lower left panel, a
simulated ensemble of X-ray binaries was produced (upper left
panel). This was compared with a sample of SDSS quasars and low
luminosityt AGN (upper right panel) and a striking similarity
revealed. The lower right panel offers an explanation for the
similarities between the different classes of object.}
\label{SDSS}
\end{figure}

\section{Beware of cheap imitations (or: Neutron stars can do it too)}

It has already been known for some years that neutron star X-ray
binaries (NSXRBs) are also radio sources. Migliari \& Fender (2006)
present a review of such observations, including cases where the radio
emission has been resolved into jets, and compare them with those from
black hole X-ray binaries. It is clear that at least at high accretion
rates, NSXRBs can also produce relativistic jets which would not look
out of place in a black hole system. Furthermore, as already noted,
K\"ording, Fender \& Migliari (2006) argue that, within a factor of a
few, NSXRB jets are as powerful as those from BHXRBs for a given
accretion rate (in hard states).

\begin{figure}
\centerline{\epsfig{file=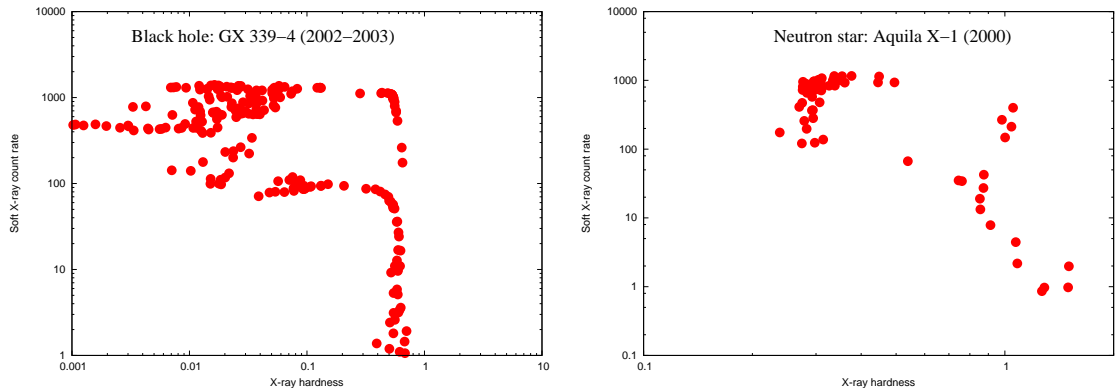, angle=0, width=15cm}}
\caption{X-ray hardness intensity diagrams for the two X-ray binaries
in outburst. The left panel is the 2002-2003 outburst of the black
hole X-ray binary GX 339-4 (Homan \& Belloni 2005). The right panel is
the 2000 outburst of the neutron star X-ray binary Aquila X-1 (Maitra
\& Bailyn 2003). The similarity in the pattern of outburst, notably
the hysteretical state changes, is clear. Note that the X-ray colours
are defined slightly differently for the two sources.}
\label{aqlx1}
\end{figure}

What about the {\em patterns} of outbursts -- perhaps those could be
used as a discriminant ? It seems not: Maitra \& Bailyn (2003) present
the HID for an outburst of the neutron star X-ray transient Aql
X-1. The HID, when plotted on similar axes to those of the black holes
(Fig 7) is striking similar (see also van der Klis 2006). This
indicates that the pattern of the outburst is not dominated by any
properties specific to black holes such as a static limit or event
horizon, but rather by properties common to both classes of object,
such as disc instabilities and accretion into a deep gravitational
potential. So while there are some patterns of behaviour -- e.g. the Z
sources -- which seem to be unique to NSXRBs, overall it does not seem
that we can attribute any key property of relativistic jet formation,
or its coupling to accretion, as arising due to some property unique
to black holes.

\section{Conclusions}

The scaling of black hole accretion with mass, from X-ray binaries to
Quasars, has evolved from a clear concept and idea in the 1970s, to a
something with a sound observational footing in the 1990s, and
eventually to a field with measured quantitative scalings with
predictive power in the 2000s. So what do we do with these scaling relations ?

The stated goal, beyond testing the hypothetical simplicity of black
hole accretion, has often been that we can use the relatively rapid
temporal evolution of black hole X-ray binaries to understand
something about the life cycles of AGN. Happily the importance of this
potential insight has only grown in recent years with the growing
acceptance that feedback from AGN directly shapes the evolution of the
host galaxy (e.g Best et al. 2005 and references therein; see also
Silk \& Rees 1998) and in some cases, clusters in which the AGN are
embedded. There are already some examples of linkage between the
fields. Based on knowledge of the relation of jet production and mass
accretion rate (caveat state hysteresis at high accretion rates), one
can for example take AGN luminosity functions and use them to estimate
the total amount of kinetic feedback associated with the accretion
process (see Merloni \& Heinz (2007) and K\"ording, Jester \& Fender
(2007)).

Finally it is of interest to note that as the Universe continues to
evolve, we will move towards a time in which the release of
gravitational energy is dominated by jets, rather than radiation, as
black holes accrete at lower and lower mean Eddington ratios. The
future is not bright.

\section*{Acknowledgments}

RPF would like to acknowledge useful conversation with Annalisa
Celotti, Stephane Corbel, Heino Falcke, Elena Gallo, Jeroen Homan,
Sebastian Jester, Christian Kaiser, Tom Maccarone, Sera Markoff, Dave
Meier, Simone Migliari and many others both at the Como workshop and
further back in the mists of time.

\end{document}